\documentclass[aps,prb,twocolumn,showpacs,a4paper,amsmath,amssymb,
byrevtex,superscriptaddress]{revtex4}

\usepackage[latin1]{inputenc}
\usepackage{graphicx}
\usepackage{dcolumn}
\usepackage{bm}
\usepackage{hyperref}
\usepackage{times}

\begin{document}

%\preprint{ffuov/02-01}

\title{Singlet-triplet filtering and entanglement in a quantum dot structure}

\author{G. Giavaras}
\affiliation{Department of Physics, Lancaster University,
Lancaster LA14YB, England} \affiliation{QinetiQ, St.Andrews Road,
Malvern WR143PS, England}
\author{J. H. Jefferson}
\affiliation{QinetiQ, St.Andrews Road, Malvern WR143PS, England}
\author{M. Fearn}
\affiliation{QinetiQ, St.Andrews Road, Malvern WR143PS, England}
\author{C. J. Lambert}
\affiliation{Department of Physics, Lancaster University,
Lancaster LA14YB, England}

\date{\today}

\begin{abstract}
We consider two interacting electrons in a semiconductor quantum
dot structure which consists of a small dot within a larger dot,
and demonstrate a singlet-triplet filtering mechanism which
involves spin-dependent resonances and can generate entanglement.
By studying the exact time evolution of singlet and triplet states
we show how the degree of both filtering and spin entanglement can
be tuned using a time-dependent gate voltage.
\end{abstract}

%\pacs{}

\maketitle

In semiconductor quantum dots which can be formed for example at
the interface of a gated GaAs/AlGaAs heterostructure, we can trap
electrons and control their number precisely down to
one\cite{kouwen1,tarucha,kouwen2,wiel}. The fact that the
electrons in the dots are well-confined within a region of some
nanometers and they are therefore isolated from the remaining
environment of the host material, offers not only the opportunity
to explore atomic-like effects and electron
correlations\cite{tarucha,kouwen2,cobden,moriyama} in the solid
state, but also to consider the quantum dots as promising
candidates for nanoelectronic applications. For instance the
proposal for realisation of spin and charge qubits in double
quantum dots\cite{loss} has attracted a lot of interest both
theoretically\cite{burkard,schliemann,hu,wisconsin,zhang} and
experimentally \cite{koppens,hanson1,hayashi,petta} and the
demonstration of entanglement, which is a necessary ingredient for
two-qubit gates, is one of the main goals. Of course many other
systems which include quantum dots make use of the Loss and
DiVincenzo\cite{loss} exchange-energy mechanism based on the
Coulomb interaction to generate spin-entanglement, such as for
example flying qubits in surface acoustic wave structures
\cite{barnes} and static-flying qubits in quantum
wires\cite{jefferson,giavaras}. The singlet-triplet qubit in a
double quantum dot is also a promising candidate for quantum
computation. The two-electron singlet-triplet ($S_{z}=0$) states
constitute the two-level quantum system and as has been shown this
scheme, which is also under experimental
investigation\cite{petta}, is efficient for universal quantum
gates \cite{hanson2,taylor}.

In this paper we propose and investigate a semiconductor dot
structure which can induce a mechanism to filter singlet-triplet
states and thus to generate maximal spin entanglement between two
electrons. The operation is controlled by a time-dependent gate
voltage which can tune the degree of filtering and entanglement.

\begin{figure}
\begin{center}
\includegraphics[width=7.3cm,height=5cm]{dot.eps}
\includegraphics[width=5.90cm,height=5cm]{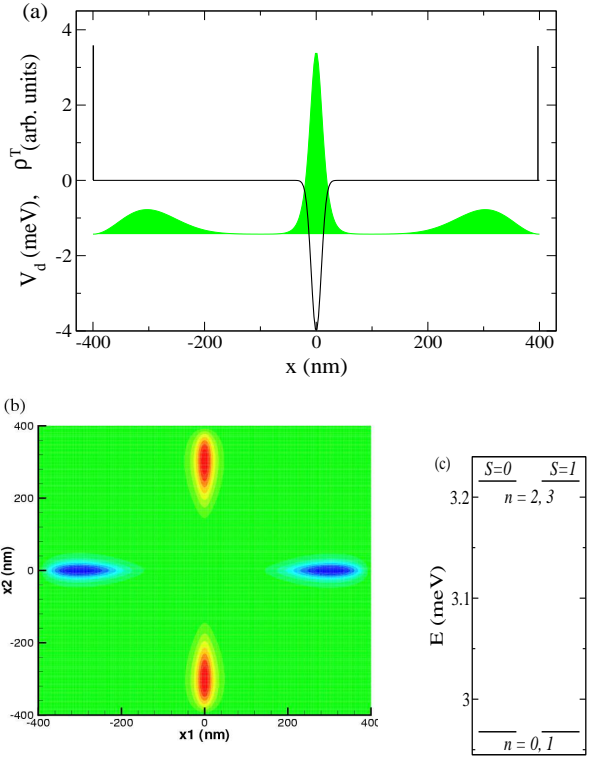}
\includegraphics[width=2.085cm,height=4cm]{ene.eps}
\caption{(Color online) (a) Quantum dot confining potential and
effective charge density (in arbitrary units) of the lowest
triplet eigenstate. (b) Contour plot of the lowest triplet
eigenstate. (c) The two-lowest singlet and triplet
pairs.}\label{dot}
\end{center}
\end{figure}

To be specific, we choose material parameters for GaAs and
consider the quasi-one-dimensional quantum dot structure which is
shown in Fig.~\ref{dot}(a), loaded with two electrons. It consists
of a small open dot which is formed in the centre of a much larger
dot. The small dot is such that it can bind only one electron and
this means that the Coulomb interaction forces the second electron
to occupy the region of the large dot. The latter has width
$L=800$ nm $\sim80a^{*}_{B}$, where $a^{*}_{B}$ is the effective
Bohr radius for GaAs. In this regime the two electrons are in the
strong correlation regime \cite{bryant,creffield} i.e. the Coulomb
interaction dominates over the kinetic energy and in the ground
state the two electrons are localised in regions for which the
electrostatic repulsion is minimised. Note that this is the case
for both singlet and triplet states which give virtually the same
electron distribution and they are separated with a very small
antiferromagnetic exchange energy $J$.

To explicitly demonstrate and analyse these effects we have
studied the two-electron problem with exact
diagonalisation\cite{note1} considering the parabolic-band
Hamiltonian
\begin{equation}
H=\sum_{i=1,2}\left[
-\frac{\hbar^{2}}{2m^{*}}\frac{\partial^{2}}{\partial
x^{2}_{i}}+V_{d}(x_{i})\right]+V_{c}(x_{1},x_{2})\label{hamilt},
\end{equation}
where $m^{*}=0.067m_{o}$ is the effective mass of the electrons
for GaAs. We have modelled the quantum dot with the Gaussian
confining potential $V_{d}(x)=-V_{o}\exp(-x^{2}/2l^{2}_{o})$, for
-400 nm $\leq$ x $\leq$ 400 nm and $V_{d}(x)=\infty$
otherwise\cite{note2}. The parameters $V_{o}, l_{o}$ determine the
depth and width of the small dot respectively and they are chosen
to give only a single bound energy level $\sim$ -1.5 meV. The
Coulomb term is given by $V_{c}(x_{1},x_{2})=q^{2}/4\pi \epsilon
_{r}\epsilon _{o} r$, with
$r=\sqrt{(x_{1}-x_{2})^{2}+\lambda_{c}^{2}}$ and $\epsilon
_{r}=13$ the relative permittivity in GaAs. This simplified form
of the Coulomb interaction assumes that all excitations take place
in the $x$-direction, whereas in the $y$, $z$-directions the
electrons occupy at all times the lowest transverse modes. To
satisfy this assumption we choose for all the calculations
$\lambda_{c}=20$ nm, which gives a physical confinement length in
the transverse $y$, $z$ directions much smaller than that in $x$.
The two-electron time-independent Schr\"odinger equation is solved
numerically by the configuration interaction method. The
low-energy eigenstates consist of two closely spaced singlets and
two triplets with a somewhat larger energy gap to higher-lying
states. Fig.~\ref{dot}(b) shows the lowest triplet eigenstate
$\Phi^{T}_{o}(x_{1},x_{2})$ (antisymmetric) and Fig.~\ref{dot}(c)
typical energy levels. The corresponding effective charge density
of $\Phi^{T}_{o}(x_{1},x_{2})$ which is derived by integrating the
two-electron charge density over the spatial coordinates of one of
the two electrons i.e.
$\rho^{T}(x)=2q\int\vert\Phi^{T}_{o}(x,x^{'})\vert^{2}dx^{'}$ is
shown in Fig.~\ref{dot}(a). For this distribution one electron is
bound in the small dot, whereas the second electron is at the left
and right corners with equal probability due to the symmetric
potential. The charge densities associated with all states in the
low-lying manifold are in fact virtually identical to that shown
in Fig.~\ref{dot}(a).

This can be understood from a simple Heitler-London picture using
one electron states derived from a Hartree approximation in which
we replace the trapped electron in the small dot with an effective
Hartree potential $V_{H}(x)=\int
V_{c}(x,x^{'})\vert\psi_{c}(x^{'})\vert^{2}dx^{'}$, where
$\psi_{c}(x)$ describes the single electron state of the small
dot. The second electron feels the approximate potential
$V_{H}(x)+V_{d}(x)$, which for the lowest two-electron states of
interest acts as a double-well potential. The two lowest
single-electron states $\psi_{-}(x)$, $\psi_{+}(x)$,
(bonding-antibonding) of this potential have a small energy
splitting and peak in the left and right corners. Using the one
electron Hartree states we may form the two lowest electron states
(independent of spin)
$\Phi_{o}(x_{1},x_{2})=\psi_{-}(x_{1})\psi_{c}(x_{2})$ and
$\Phi_{1}(x_{1},x_{2})=\psi_{+}(x_{1})\psi_{c}(x_{2})$ and
similarly for the next higher pair. Within the Hartree model this
means that not only the lowest singlet, triplet eigenstates have a
distribution of the form of Fig.~\ref{dot}(a), but also the first
excited eigenstates which indeed agrees with the results that we
derive from the exact diagonalisation. Note that even though the
Hartree model gives great insight into the two-electron energies
and distributions of the low-lying states, with remarkably little
effort, the symmetric and antisymmetric combinations
$\Phi^{S,A}(x_{1},x_{2})=[ \Phi_{o}(x_{1},x_{2}) \pm
\Phi_{o}(x_{2},x_{1}) ]/\sqrt{2}$ give a poor approximation to the
singlet-triplet splitting, with even the incorrect sign. This is
little improved by using more accurate one-electron states, e.g.
from Hartree-Fock or DFT, being a consequence of strong
correlations requiring accurate solutions of the two-electron
problem. However, as we discuss later, the system may be described
accurately by an extended Hubbard model.

\begin{figure}
\begin{center}
\includegraphics[width=7.5cm,height=5.5cm]{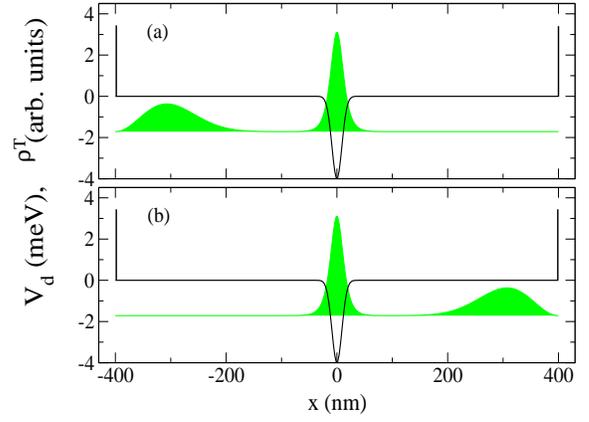}
\caption{(Color online) Effective charge density for (a) left and
(b) right triplet states. The quantum dot confining potential is
also shown.}\label{states}
\end{center}
\end{figure}

For the dynamics that we describe below it is important to
introduce left and right states which are formed by combining the
two lowest eigenstates for singlet
$\Phi^{S}_{L}=\Phi^{S}_{o}+\Phi^{S}_{1}$,
$\Phi^{S}_{R}=\Phi^{S}_{o}-\Phi^{S}_{1}$ and triplet
$\Phi^{T}_{L}=-\Phi^{T}_{o}+\Phi^{T}_{1}$,
$\Phi^{T}_{R}=\Phi^{T}_{o}+\Phi^{T}_{1}$ (unnormalised). For these
states one electron is bound in the small dot whereas the second
electron is localised to the corresponding corner as we
demonstrate in Fig.~\ref{states} for the triplet.

Although the effective charge density of the low-lying singlets
and triplets is virtually identical, the wave functions are quite
different and give rise to effective interactions between spins,
responsible for their entanglement. To demonstrate how these
states can generate spin-entanglement and give rise to
singlet-triplet filtering, we first prepare the system such that
one electron is located on the left with spin up and the other in
the central dot with spin down. A suitable non-entangled state of
this type may be written as a superposition of the left singlet
and the left $S_{z}=0$ triplet state i.e.
$\Psi_{\uparrow\downarrow}=(\Phi^{S}_{L}\chi_{\uparrow\downarrow}^{S}
+ \Phi^{T}_{L}\chi_{\uparrow\downarrow}^{T} ) /\surd2$, with the
spin eigenstates
$\chi^{S/T}_{\uparrow\downarrow}(1,2)=[\chi_{\uparrow}(1)\chi_{\downarrow}
(2)\mp\chi_{\downarrow}(1)\chi_{\uparrow}(2)]/\surd{2}$. An
approximation to this state may be obtained in principle by
applying a small source-drain bias, $V_{sd}$, across the quantum
dot structure and allowing the system to relax to its ground state
(singlet). The spins may then be initialised using magnetic fields
and a microwave pulse, as suggested for scalable qubit
arrays\cite{loss,koppens}. If $V_{sd}$ is removed, then
oscillations (in a similar fashion to a double dot system) and
entanglement will develop with a typical time for the triplet
component of the wave function on the order of $\pi\hbar/\Delta
E^{T} \sim$ 1 $\mu$s, where $\Delta E^{T}=E^{T}_{1}-E^{T}_{o}$ is
the energy splitting of the two lowest triplet eigenstates when
$V_{sd}$=0 (similarly for the singlet). To speed-up the process we
can apply a time-dependent gate voltage that will tune in such a
way the dot potential so that to increase this energy splitting.
In this work we have modelled the gate voltage potential using the
expression $V_{g}(x,t)=-V_{p}(t)\exp(-x^{2}/2l^{2}_{g})$,
($l_{g}\sim$ 130 nm) which is driven by a triangular pulse of the
form
\begin{equation}
V_{p}(t)=\left\{
\begin{array}{c}
\frac{2V_{b}}{T_{p}}t,           \qquad    0 \leq t \leq T_{p}/2\\ \\
-\frac{2V_{b}}{T_{p}}t+2V_{b},   \qquad    T_{p}/2\leq t \leq T_{p},
\end{array}
\right.\label{pulse}
\end{equation}
where $T_{p}$ is the period of the pulse and $V_{b}$ the maximum
gate voltage. We have studied the case for which the gating rate
$\alpha=\vert dV_{p}/dt\vert$ is such that to a good approximation
only the two lowest singlet and two lowest triplet states are
involved in the dynamics. During the first half of the cycle (0
$\leq$ t $\leq T_{p}/2$) the effect of the voltage is to decrease
the effective width of the dot thereby increasing the energy
splitting and the interaction, whereas during the second half of
the cycle ($T_{p}/2 \le$ t $\leq T_{p}$ ) the process is reversed.

\begin{figure}
\begin{center}
\includegraphics[width=7.7cm,height=7.0cm]{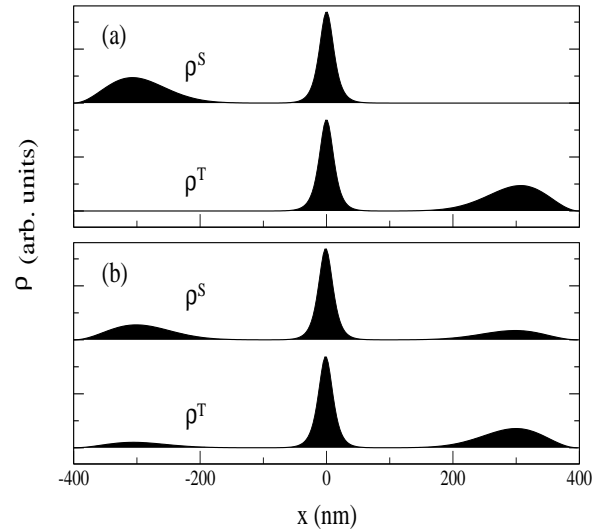}
\caption{Effective charge density for singlet and triplet
components at the final time for (a) a case in the filtering
regime and (b) the most general case.}\label{evo}
\end{center}
\end{figure}

The dynamics of the two electrons is governed by the
time-dependent Schr\"odinger equation with the Hamiltonian (1) and
for a total potential $V_{d}(x)+V_{g}(x,t)$ which is the sum of
the dot confining potential and the time-dependent gate potential.
For $t>0$, the spin eigenstates are unchanged for singlet or
triplet components, whereas the evolution of the corresponding
orbital states is given directly by the solution of the
time-dependent Schr\"odinger equation which is implemented
numerically using the staggered-time algorithm proposed by
Visscher\cite{visscher}. The time evolution of the initial state,
$\Psi_{\uparrow\downarrow}$, is then determined by adding the
separately determined singlet and triplet components. In
Fig.~\ref{evo} we show two examples of the final electron
distribution for singlet and triplet components. Specifically, in
Fig.~\ref{evo}(a) the singlet corresponds to a good approximation
to the right state, whereas the triplet to the left state. This is
the ideal filtering regime (together with the opposite limit) in
which at the final time the two components occupy different
spatial regions. Note that in this regime the two electrons are
fully entangled provided the measurement domain is restricted
either to the left or right region, detecting the singlet or the
$S_{z}=0$ triplet state respectively, which are fully entangled
states. Performing an additional measurement but initialising the
electron spins to a purely triplet state (either of $S_{z}=\pm1$)
reveals the occupation region of the triplet state hence offering
a way to discriminate between singlet and triplet components. Note
that in the most general case which is shown in Fig.~\ref{evo}(b)
the final two-electron state is a superposition of left and right
states for both singlet and triplet components. Clearly, a
measurement with restriction to either the left domain or the
right domain would probe a partially entangled state.

\begin{figure}
\begin{center}
\includegraphics[width=7.7cm,height=6.cm]{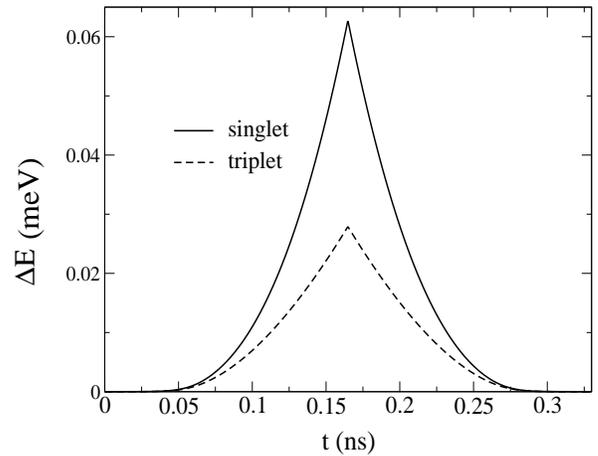}
\caption{Energy splitting of the two-lowest levels for singlet and
triplet components, as a function of time for one cycle in the
filtering regime.}\label{splitting}
\end{center}
\end{figure}

To understand the dynamics and the difference between singlet and
triplet components we can write the time-dependence, for example
of the singlet, using the corresponding two lowest energy states
at all times i.e.
$\Phi^{S}=C^{S}_{L}\Phi^{S}_{L}+C^{S}_{R}\Phi^{S}_{R}$ in the
left-right basis with the amplitudes
$C^{S}_{L}=e^{-i\varphi^{S}}\cos\omega^{S}$ and
$C^{S}_{R}=ie^{-i\varphi^{S}}\sin\omega^{S}$. The parameters
$\varphi^{S}=\int_{0}^{t}[E^{S}_{1}(t^{'})+E^{S}_{o}(t^{'})]dt^{'}/2\hbar$
and
$\omega^{S}=\int_{0}^{t}[E^{S}_{1}(t^{'})-E^{S}_{o}(t^{'})]dt^{'}/2\hbar$
depend on the two lowest energy levels and change with time. Note
that $\omega^{S}$ determines the amount of filtering because it
determines the period of the oscillations between left and right
states, whereas both $\varphi^{S}$ and $\omega^{S}$ determine the
degree of entanglement as we demonstrate below. In
Fig.~\ref{splitting} we show the energy splitting of the two
lowest levels (the energies are calculated from instantaneous
solutions) as a function of time for one cycle in the filtering
regime which gives fully entangled electrons in time
$T_{p}\sim$0.3 ns (for $V_{b}\sim$2.8 meV). We see that the
splitting is larger for the singlet and this means that
$\omega^{S}>\omega^{T}$ i.e. the singlet component oscillates
faster than the triplet. We can understand this effect with a
Hartree approximation: because the small dot has only one bound
energy level, $\varepsilon_{c}$, the tunneling time (the time that
the electron in the left corner needs in order to move to the
right corner) for the triplet is longer due to the Pauli blocking.
Within the Hartree model the electron in the large dot needs to
tunnel through an effective double barrier due to the Coulomb
repulsion from the trapped electron in the small dot, which has
only one resonance energy level $\sim\varepsilon_{c}+U_{c}$ (with
$U_{c}$ the Coulomb energy when both electrons occupy
$\varepsilon_{c}$) and corresponds to a singlet state. For a
triplet resonance to exist the small dot needs to have at least
two bound levels due to the Pauli principle. It is worth noting
that the efficiency of this singlet-triplet filtering mechanism
and the induced entanglement has been also demonstrated in
scattering problems between a static and a flying
qubit\cite{jefferson,giavaras}.

\begin{figure}
\begin{center}
\includegraphics[width=7.7cm,height=7.4cm]{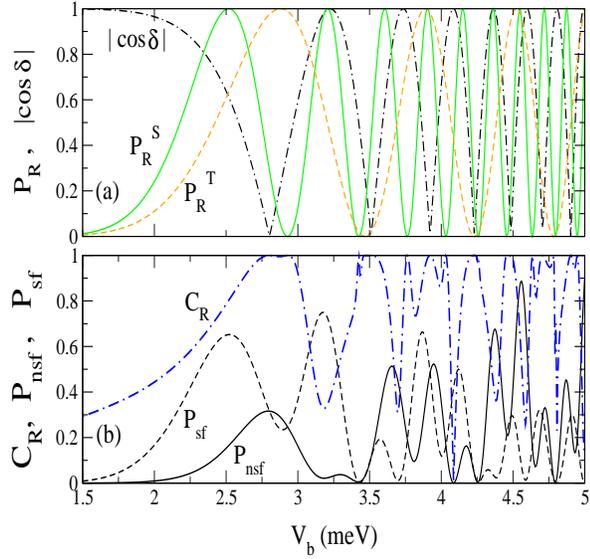}
\caption{(Color online) Dependence (at the final time), as a
function of maximum gate voltage for fixed gating rate, of: (a)
absolute value of $\cos\delta$ and right probabilities for singlet
and triplet components; (b) concurrence, spin-flip and non
spin-flip probabilities calculated for the right
region.}\label{conc1}
\end{center}
\end{figure}

To quantify the degree of entanglement which arises as a
consequence of the interaction when the electrons are close
together we have calculated concurrence \cite{wooter} at the final
time, from spin-flip, $P_{sf}$, and non spin-flip, $P_{nsf}$,
probabilities. For example at the right region the amplitudes
for these probabilities are
\begin{equation}
C_{nsf}=\frac{C^{T}_{R}+C^{S}_{R}}{2}, \quad
C_{sf}=\frac{C^{T}_{R}-C^{S}_{R}}{2},
\end{equation}
which are derived by writing the corresponding final state
$\Psi_{\uparrow\downarrow}(t_{f})$ in the basis of spin
eigenstates. Concurrence at the right region is then given by
\begin{equation}
C_{R}=\frac{2(P_{nsf}P_{sf})^{1/2}}{P_{sf}+P_{nsf}},\label{3}
\end{equation}
with $P_{nsf}=\vert C_{nsf} \vert^{2}$ and $P_{sf}=\vert C_{sf} \vert^{2}$,
thus
\begin{equation}
C_{R}=\frac{[(P^{S}_{R}+P^{T}_{R})^{2}-4
P^{S}_{R}P^{T}_{R}\cos^{2}\delta]^{1/2}}  {P^{S}_{R}+P^{T}_{R}},\label{4}
\end{equation}
where we have set $P^{S}_{R}=\vert
C^{S}_{R}\vert^{2}=\sin^{2}\omega^{S}$ and $P^{T}_{R}=\vert
C^{T}_{R}\vert^{2}=\sin^{2}\omega^{T}$ and the relative phase
$\delta=\varphi^{S}-\varphi^{T}$. A similar expression can be
derived for the left and even for the total region. Note that by
definition $0\leq C_{R} \leq 1$ where the limit $C_{R}=0$
corresponds to unentangled electrons, whereas the limit $C_{R}=1$
to fully entangled electrons. The measurement in the right region
is meaningful only when $P^{S}_{R}\neq0$, and/or $P^{T}_{R}\neq0$
i.e. when the right state is occupied for singlet and/or triplet.
We see from Eq.~(\ref{3}) that $C_{R}=1$ when $P_{sf}$=$P_{nsf}$
or equivalently from Eq.~(\ref{4}) when $\cos\delta=0$ and/or when
$P^{S}_{R}=0$ and simultaneously $P^{T}_{R}\neq0$ or vice versa.

\begin{figure}
\begin{center}
\includegraphics[width=8.5cm,height=7.4cm]{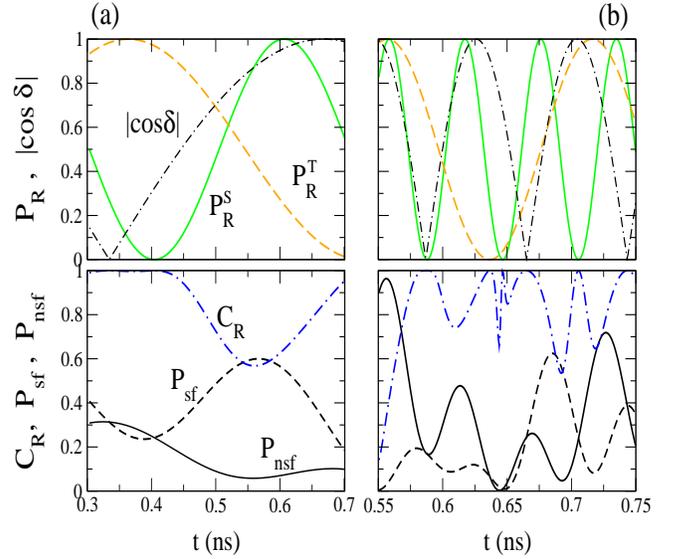}
\caption{(Color online) Dependence (at the final time) of quantities
in Fig.~\ref{conc1} as a function of pulse duration time for two
different maximum gate voltages. (a) $V_{b}=$2.75 meV and (b)
$V_{b}=$5 meV.}\label{conc2}
\end{center}
\end{figure}

In Fig.~\ref{conc1} we present the dependence of probabilities,
phase shift and concurrence as a function of maximum gate voltage
$V_{b}$ for a fixed gating rate $\alpha \sim$ 18 meV/ns which
ensures that to a good approximation only the two lowest
eigenstates for both singlet and triplet components are involved
into the dynamics. Figure~\ref{conc1}(a) shows the probabilities
$P^{S}_{R}$, $P^{T}_{R}$ and the absolute magnitude of the
quantity $\cos\delta$ calculated at the final time, versus
$V_{b}$. We see that the probabilities $P^{S}_{R}$, $P^{T}_{R}$ do
not oscillate for small gates voltages $V_{b}\lesssim2.5$ meV (but
rather they increase) because the energy splitting of the
two-lowest levels in this limit remains relatively small and
therefore the tunneling time long. Further increase of the gate
voltage induces well-defined oscillations. In particular the
probability of the singlet component oscillates faster than the
triplet because the energy splitting which determines the
frequency of the oscillations is larger for the singlet. Note also
that the frequency of oscillations increases with gate voltage for
both components following the increase of the energy splitting.
Figure \ref{conc1}(b) shows the concurrence $C_{R}$ and the
spin-flip $P_{sf}$ and non spin-flip $P_{nsf}$ probabilities
calculated at the final time. When these probabilities are equal
the entanglement is maximum whereas when one of them is zero the
electrons remain unentangled. For a gate voltage $V_{b}\sim$2.8
meV we have an example of the ideal filtering regime for which
$P^{S}_{R}\sim$0, $P^{T}_{R}\sim$1 and $P_{sf}=P_{nsf}$ giving
$C_{R}$=1, since only a triplet state occupies the right region.
The generic condition for ideal filtering is when
$\sin\omega^{S}$=1 and simultaneously $\sin\omega^{T}$=0 or vice
versa. Other filtering cases (non-ideal) which give $C_{R}=$1
occur when one of the components is zero, say $P^{S}_{R}=$0 and
the other is non-zero $P^{T}_{R}\neq$0, but not however equal to
one. Finally an interesting limit is when $P^{S}_{R}=P^{T}_{R}$
for which the concurrence reduces to the simple expression
$C_{R}=\vert\sin\delta\vert$ which depends only on the relative
phase between singlet and triplet components.

Figure \ref{conc2} presents the dependence of various quantities
as a function of the pulse duration time for two different gate
voltages. As we see by comparing Figs.~\ref{conc2}(a) and
\ref{conc2}(b) the period of the oscillations can be controlled
with the value of the maximum gate voltage with the larger $V_{b}$
inducing faster oscillations due to the larger energy splitting.

\begin{figure}[t]
\begin{center}
\includegraphics[width=7.7cm,height=4.4cm]{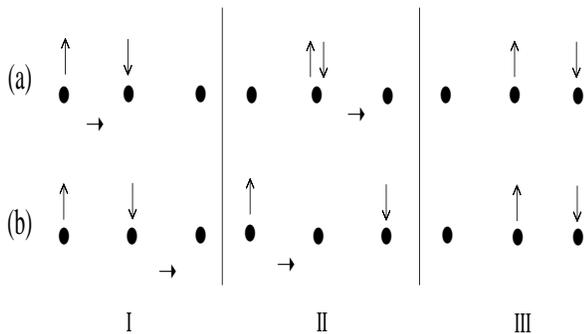}
\caption{Illustration of the two-electron spin dynamics in a
three-step process I, II, III. The process (a) is allowed only for
singlets whereas the process (b) is allowed for both singlets and
triplets.}\label{hubb}
\end{center}
\end{figure}

Finally, we point out that the low-lying spin states relevant to
the dynamics of this entangler which involve virtually identical
charge distributions at the left, centre and right of the dot
structure suggest that this charge-spin system may be modelled by
a three-site Hubbard model of the form
\begin{equation}
\begin{split}
H_{eff}= & \varepsilon(n_{L}+n_{R})+\varepsilon_{c}n_{C} \\
         &
+\gamma\sum_{\sigma,i=L,R}(C^{\dagger}_{i,\sigma}C_{C,\sigma}+\text{h.c.}) \\
         & +U( n_{L,\uparrow}n_{L,\downarrow} + n_{R,\uparrow}n_{R,\downarrow})
+ U_{c}n_{C,\uparrow}n_{C,\downarrow}\\  & + V(n_{L}n_{C} +n_{C}n_{R}), \\
\end{split}\label{hubba}
\end{equation}
where $n_{i}=\sum_{\sigma}n_{i,\sigma}=\sum_{\sigma}
C^{\dagger}_{i,\sigma} C_{i,\sigma}$, $i=L,R,C$ and
$C^{\dagger}_{i,\sigma}( C_{i,\sigma})$ creates (destroys) an
electron on site $i$ with spin-$\sigma$. The on-site orbital
energy for the left and right sites is $\varepsilon$ and for the
central site is $\varepsilon_{c}$. The on-site Coulomb energy is
denoted by $U$ for the left and right sites and by $U_{c}$ for the
central site. The nearest site Coulomb energy is $V$ and finally
$\gamma$ expresses the hopping between nearest sites. Note that
all the physical parameters ($\varepsilon$, $\varepsilon_{c}$,
$U$, $U_{c}$, $V$, $\gamma$) may be estimated from the Hartree
approximation described earlier. Working within the restricted
subspace for which the two-electron basis states consist of six
singlets and three triplets, we can extract the correct energy
splitting of the two-lowest eigenstates (for both singlet and
triplet), the corresponding eigenstates and the antiferromagnetic
exchange energy. Solution of this time-dependent model
Eq.~(\ref{hubba}) does indeed show qualitatively the same
behaviour as the original continuous problem  but with
considerable saving in computer time once the time-dependent
parameters are known. The analysis based on the effective
Hamiltonian Eq.~(\ref{hubba}) indicates that the physical
behaviour that we have demonstrated can also be realised in a
triple quantum dot structure. More importantly, the Hubbard model
gives insight into the behaviour of the system since it is readily
mapped onto an effective charge-spin model for the low-lying
manifold of two singlets and two triplets. In particular, for
singlets there are two processes by which an electron may tunnel
from left to right, as shown in Fig.~\ref{hubb}. The process (a),
that is only allowed for singlets, has amplitude
$\Delta_{1}=\gamma^{2}/(U-V+\varepsilon_{c}-\varepsilon)\sim J$
where $J$ is the Heisenberg exchange energy between an electron
spin on the central site and one on either the left or the right
site. On the other hand, process (b) is valid for both singlets
and triplets and has amplitude
$\Delta_{2}=\gamma^{2}/(\varepsilon-\varepsilon_{c}-V)$. The
relative rate of tunneling for singlet and triplet is thus tuned
by the gate by changing $\varepsilon_{c}$, since making
$\varepsilon_{c}$ more negative the energy denominator in
$\Delta_{1}$ decreases whereas that of $\Delta_{2}$ increases.

In summary we have presented a quantum dot structure and described
the two-electron distribution for the lowest singlet and triplet
states. By studying the electron dynamics due to a time-dependent
gate voltage we showed how we can induce a singlet-triplet
filtering based on the Pauli blocking effect for the triplet
state. This can generate full spin-entanglement within the order
of $\sim$0.3 ns. Both the degree of filtering and entanglement can
be efficiently controlled with the gating rate and the maximum
applied gate voltage.

G.G. thanks UK EPSRC for funding. This work was supported by the
UK Ministry of Defence.


\begin{thebibliography}{99}

\bibitem{kouwen1}
L. P. Kouwenhoven, T. H. Oosterkamp, M. W. S. Danoeastro, M. Eto, D. G. Austing,
T. Honda, and S. Tarucha, Science \textbf{278}, 1788 (1997).

\bibitem{tarucha}
S. Tarucha, D. G. Austing, T. Honda, R. J. van der Hage, and L. P. Kouwenhoven,
Phys. Rev. Lett. \textbf{77}, 3613 (1996).

\bibitem{kouwen2}
L. P. Kouwenhoven, D. G. Austing, and S. Tarucha, Rep. Prog. Phys. \textbf{64},
701 (2001).

\bibitem{wiel}
W. G. van der Wiel, S. De Franceschi, J. M. Elzerman, T. Fujisawa, S. Tarucha,
and L. P. Kouwenhoven, Rev. Mod. Phys. \textbf{75}, 1 (2003).


\bibitem{cobden}
D. H. Cobden and J. Nygard, Phys. Rev. Lett. \textbf{89}, 046803
(2002).

\bibitem{moriyama}
S. Moriyama, T. Fuse, M. Suzuki, Y. Aoyagi, and K. Ishibashi, Phys. Rev. Lett.
\textbf{94}, 186806 (2005).





\bibitem{loss}
D. Loss and D. P. DiVincenzo, Phys. Rev. A \textbf{57}, 120 (1998).

\bibitem{burkard}
G. Burkard, D. Loss, and D. P. DiVincenzo, Phys. Rev. B \textbf{59}, 2070
(1999).

\bibitem{schliemann}
J. Schliemann, D. Loss, and A. H. MacDonald, Phys. Rev. B \textbf{63}, 085311
(2001).

\bibitem{hu}
X. Hu and S. Das. Sarma, Phys. Rev. B \textbf{61}, 062301 (2000).

\bibitem{wisconsin}
M. Friesen, P. Rugheimer, D. E. Savage, M. G. Lagally, D. W. van der Weide, R.
Joynt, and M. A. Eriksson, Phys. Rev. B \textbf{67}, 121301(R) (2003).

\bibitem{zhang}
P. Zhang, Q. K. Xue, X. G. Zhao, and X. C. Xie, Phys. Rev. A \textbf{66}, 022117
(2002).



\bibitem{koppens}
F. H. L. Koppens, C. Buizert, K. J. Tielrooij, I. T. Vink, K. C. Nowack. T.
Meunier, L. P. Kouwenhoven, and L. M. K. Vandersypen, Nature \textbf{442}, 766
(2006).

\bibitem{hanson1}
R. Hanson, L. H. Willems van Beveren, I. T. Vink, J. M. Elzerman, W. J. M.
Naber, F. H. L. Koppens, L. P. Kouwenhoven, and L. M. K. Vandersypen, Phys.
Rev. Lett. \textbf{94} 196802 (2005).

\bibitem{hayashi}
T. Hayashi, F. Fujisawa, H. D. Cheong, Y. H. Jeong, and Y. Hirayama, Phys. Rev.
Lett. \textbf{91}, 226804 (2003).

\bibitem{petta}
J. R. Petta, A. C. Johnson, J. M. Taylor, E. A. Laird, A. Yacoby, M. D. Lukin,
C. M. Marcus, M. P. Hanson, A. C. Gossard, Science \textbf{309}, 2180 (2005).


\bibitem{barnes}
C. H. W. Barnes, J. M. Shilton, and A. M. Robinson, Phys. Rev. B
\textbf{62}, 8410 (2000).

\bibitem{jefferson}
J. H. Jefferson, A. Ram$\check{s}$ak, and T. Rejec, Europhys. Lett. \textbf{75},
764
(2006).

\bibitem{giavaras}
G. Giavaras, J. H. Jefferson, A. Ram$\check{s}$ak, T. P. Spiller, and C. J.
Lambert, Phys. Rev. B \textit{in press}, cond-mat/0610168.


\bibitem{hanson2}
R. Hanson and G. Burkard, cond-mat/0605576 (2006).

\bibitem{taylor}
J. M. Taylor, H.-A.Engel, W. D\"ur, A. Yacoby, C. M. Marcus, P. Zoller,
and M. D. Lukin, Nature Physics \textbf{1}, 177 (2005).


\bibitem{bryant}
G. W. Bryant, Phys. Rev. Lett. \textbf{59}, 1140  (1987).

\bibitem{creffield}
C. E. Creffield, W. H\"ausler, J. H. Jefferson, and S. Sarkar, Phys. Rev. B
\textbf{59},
10719 (1999).

\bibitem{note1}
For two-electrons
$\Psi(x_{1},\sigma_{1},x_{2},\sigma_{2})=\Phi(x_{1},x_{2})\chi(\sigma_{1},
\sigma_{2})$ with $\Phi$ symmetric (antisymmetric) for singlet (triplet) under
electron exchange. Because the Hamiltonian does not contain any spin dependent
term we can consider only the spatial components.

\bibitem{note2}
We have used this potential form mainly for simplicity. Note however that the
main results that we present do not depend qualitatively on this particular
choice and they can be easily extended to more realistic potential profiles.


\bibitem{visscher}
P. B. Visscher, Comput. Phys. \textbf{5}, 596 (1991).

\bibitem{wooter}
W. K. Wootters, Phys. Rev. Lett. \textbf{80}, 2245 (1998).








\end{thebibliography}
\end{document}